\documentclass[twocolumn,twocolappendix,trackchanges]{aastex63}
\usepackage{bm}
\usepackage{braket}
\usepackage{amsmath}
\usepackage{subeqnarray}
\usepackage{here}
\usepackage{multirow}
\usepackage{makecell}
\usepackage{tabularx}
\usepackage{threeparttable}
\usepackage{lipsum}

\received{24-Sep-2024}
\accepted{26-Nov-2024}
\submitjournal{ApJ}
\shorttitle{}
\shortauthors{Yoshihisa et al.}
\begin{document}

\title{Conditions for Solar Prominence Formation Triggered by Single Localized Heating}

\correspondingauthor{Takero Yoshihisa}
\email{yoshihisa@kusastro.kyoto-u.ac.jp}

\author[0000-0001-5457-4999]{Takero Yoshihisa}
\affiliation{\rm{Astronomical Observatory, Kyoto University, Sakyo-ku, Kyoto 606-8502, Japan}}

\author[0000-0001-5457-4999]{Takaaki Yokoyama}
\affiliation{\rm{Astronomical Observatory, Kyoto University, Sakyo-ku, Kyoto 606-8502, Japan}}

\author[0000-0001-5457-4999]{Takafumi Kaneko}
\affiliation{\rm{Niigata University, 8050 Ikarashi 2-no-cho, Nishi-ku, Niigata 950-2181, Japan}}

\begin{abstract}

We performed numerical simulations to study mechanisms of solar prominence formation triggered by a single heating event.
In the widely accepted ``chromospheric-evaporation condensation" model, localized heating at footpoints of a coronal loop drives plasma evaporation and eventually triggers condensation.
The occurrence of condensation is strongly influenced by the characteristics of the heating.
Various theoretical studies have been conducted along one-dimensional field lines with quasi-steady localized heating.
The quasi-steady heating is regarded as the collection of multiple heating events among multiple strands constituting a coronal loop.
However, it is reasonable to consider a single heating event along a single field line as an elemental unit.
We investigated the condensation phenomenon triggered by a single heating event using 1.5-dimensional magnetohydrodynamic simulations.
By varying the magnitude of the localized heating rate, we explored the conditions necessary for condensation.
We found that when a heating rate approximately $\sim 10^{4}$ times greater than that of steady heating was applied, condensation occurred.
Condensation was observed when the thermal conduction efficiency in the loop became lower than the cooling efficiency, with the cooling rate significantly exceeding the heating rate.
Using the loop length $L$ and the Field length $\lambda_{\mathrm{F}}$, the condition for condensation is expressed as $\lambda_{\mathrm{F}} \lesssim L/2$ under conditions where cooling exceeds heating.
We extended the analytically derived condition for thermal non-equilibrium to a formulation based on heating amount.

\end{abstract}


\section{Introduction} \label{sec:intro}

Solar prominences and coronal rains are characterized by being one hundred times cooler and denser than the surrounding corona, with temperature $T\sim 10^{4} \ \mathrm{K}$ and electron density $n_{\mathrm{e}} \sim 10^{11} \ \mathrm{cm^{-3}}$. Although the formation mechanisms have been debated for decades, a complete understanding remains elusive.

Several observations have reported that prominences and coronal rains undergo thermal evolution from high to low temperatures over several hours during their formation \citep{liu2012first,berger2012sdo,antolin2015multi}.
This phenomenon is referred to as condensation.
Various models have been proposed to explain this process, including the ``chromospheric-evaporation" model \citep[e.g.,][]{mok1990prominence,antiochos1991model}, the ``reconnection" model \citep[e.g.,][]{linker2001magnetohydrodynamic,kaneko2015numerical}, and the ``shearing motion" model \citep{choe1992formation}.
The ``chromospheric-evaporation condensation" model is one of the most notable formation mechanisms.
In this model, evaporated plasma, driven by quasi-steady localized heating at the footpoints of magnetic loops, increases the coronal density.
This dense plasma then undergoes runaway cooling, leading to condensation.
If the heating is sufficiently concentrated at the loop footpoints, an equilibrium state cannot be established.
The loop continuously attempts to find a new equilibrium state, but no combination of temperatures, densities, and velocity allows it to achieve one \citep{klimchuk2019distinction}.
This behavior is due to thermal non-equilibrium (TNE), with chromospheric evaporation as a key factor.
Runaway cooling is attributed to thermal instability \citep[TI;][]{parker1953instability,field1965thermal}, which is a local process in the corona \citep{antolin2022multi}.

When prominences are observed at high resolution, fine structures known as threads become discerned \citep[e.g.,][]{engvold1976fine,zirker1998counter,lin2005thin}.
The neighboring structures in the corona are isolated due to the low plasma $\beta$. 
Additionally, thermal conduction primarily works along magnetic field lines.
Under these conditions, numerous numerical studies have used one-dimensional (1D) simulations along magnetic field lines to understand the formation of prominences 
\citep{antiochos1999dynamic, Antiochos2000, karpen2003constraints, karpen2005prominence,karpen2008condensation,xia2011formation,luna2012formation,huang2021unified,guo2021formation} and coronal rains \citep{karpen2001magnetic,muller2003dynamics,muller2004dynamics,mendoza2005catastrophic,antolin2010coronal,susino2010signatures,mikic2013importance,johnston2019effects,antolin2019thermal,reep2020electron,kucera2024modeling}.
Reducing the dimension in the calculations effectively resolves the large density differences between the corona and the prominence.

There are several critical physical conditions necessary for the occurrence of condensation.
In \cite{karpen2008condensation}, intermittent impulsive heating was applied at each loop footpoint, with the inter-pulse interval time ($\tau_\mathrm{int}$) varying between 0, 500, and 2000 seconds.
They demonstrated that $\tau_{\mathrm{int}}$ must be shorter than the cooling time ($\tau_{\mathrm{cool}}\sim 1,960$ s) for continuous plasma cooling to take place; otherwise, the evaporated plasma drains out of the loop \citep[see also][]{johnston2019effects}.
Heating that satisfies the condition, $\tau_{\mathrm{int}}<\tau_{\mathrm{cool}}$, is referred to as quasi-steady heating.
Another important factor is the ratio between the footpoint-localized heating rate ($Q_{\mathrm{foot}}$) and the background heating rate ($Q_{\mathrm{bg}}$).
When $Q_{\mathrm{bg}}$ is sufficiently large, condensation does not occur \citep[e.g.,][]{johnston2019effects,klimchuk2019role}, as the energy imbalance in the corona is compensated for.
Furthermore, the symmetry of the heating distribution plays a crucial role in condensation \citep{mikic2013importance,froment2018occurrence,pelouze2022role}.
\cite{mikic2013importance} found that, prior to condensation, asymmetry allows plasma to be transported to one footpoint by siphon flows. 
Regarding the occurrence of TNE, including a phenomenon called ``long-period intensity pulsations" \citep{auchere2014long,froment2015evidence}, \cite{froment2018occurrence} also highlighted the importance of the heating distribution. 
Recently, \cite{lu2024periodic} reported on a 3D simulation analysis of TNE that considers these conditions.

The authors of the aforementioned papers have assumed steady or quasi-steady footpoint-localized heating which is consistent with observations \citep[e.g.,][]{aschwanden2000evidence,aschwanden2001modeling}.
However, while it is reasonable to consider that each loop is heated individually, the superposition of multiple loops may give the appearance of intermittent heating.
When studying physical processes along a single field line, it is essential to regard a single heating event as an elemental unit. 
Here, we define a single heating as one that occurs only once, with its duration being sufficiently shorter than the cooling time.
This situation does not meet the requirement for the interval time between heating events $\tau_{\mathrm{int}}$ and cooling time $\tau_{\mathrm{cool}}$, i.e., $\tau_{\mathrm{int}}<\tau_{\mathrm{cool}}$.
A single heating does not need to occur symmetrically, and therefore may not satisfy the symmetry condition.
Such a condensation process was observed in the coronal rain example reported by \cite{kohutova2019formation}. 

From the perspective of numerical studies, \cite{reep2020electron} suggested that impulsive heating associated with electron beams cannot reproduce coronal rain, attributing the inability to trigger condensation to the short duration of the heating.
\cite{kohutova2020self} conducted a self-consistent 3D magnetohydrodynamic (MHD) simulation and demonstrated that coronal rain is produced in active region coronal loops through episodic impulsive heating.
In their study, the heating event, associated with magnetic field braiding, triggers thermal instability when its duration exceeds the cooling time.
\cite{huang2021unified} also reproduced prominence with a single impulsive heating event in the chromosphere.

Our study aims to investigate the conditions for condensation triggered by a single heating event localized at a single footpoint in a dipped loop, considering background coronal heating.
We solve the 1.5D MHD equations, which treat spatially 1D while accounting for the three components for velocity and magnetic field along the loop.
In previous studies, artificial background heating terms were applied.
However, our study accounts for energy dissipation by shock waves and Alfv$\Acute{\mathrm{e}}$n wave turbulence. 
This model enables us to reproduce a more realistic atmospheric behavior compared to earlier studies.
We incorporate localized heating at a coronal footpoint, with its duration being much shorter than cooling time.
First, we examine the capability of a single localized heating to produce condensation. 
A parameter survey was conducted to determine the required heating magnitude.
Additionally, we formulate the necessary heating amount, extending the applicability of the analytically derived condition from \cite{klimchuk2019role}.

We describe the numerical setting in Section \ref{sec:numerical setup}. In Section \ref{sec:results}, we present the results of the simulations and parameter survey. Section \ref{sec:Discussion} provides a discussion of these results, and the conclusion is given in Section \ref{sec:summary}.

\section{NUMERICAL SETUP}\label{sec:numerical setup}

To demonstrate prominence formation by a single heating event, we perform 1.5D magnetohydrodynamic (MHD) simulations that are spatially 1D while accounting for all three components of velocity and magnetic field. The equations solved include the effects of radiative cooling, thermal conduction, gravity, and energy dissipation by shock waves and Alfv$\Acute{\mathrm{e}}$n wave turbulence.
This approach builds upon previous studies on solar wind acceleration \citep{shoda2018frequency,shoda2018self,shimizu2022role} and the coronal heating problem \citep{shoda2021corona, washinoue2022effect}.

\subsection{Basic Equations and Setting}\label{sec:basic equations and setting}

\begin{figure}[!]
  \epsscale{1.1}
  \plotone{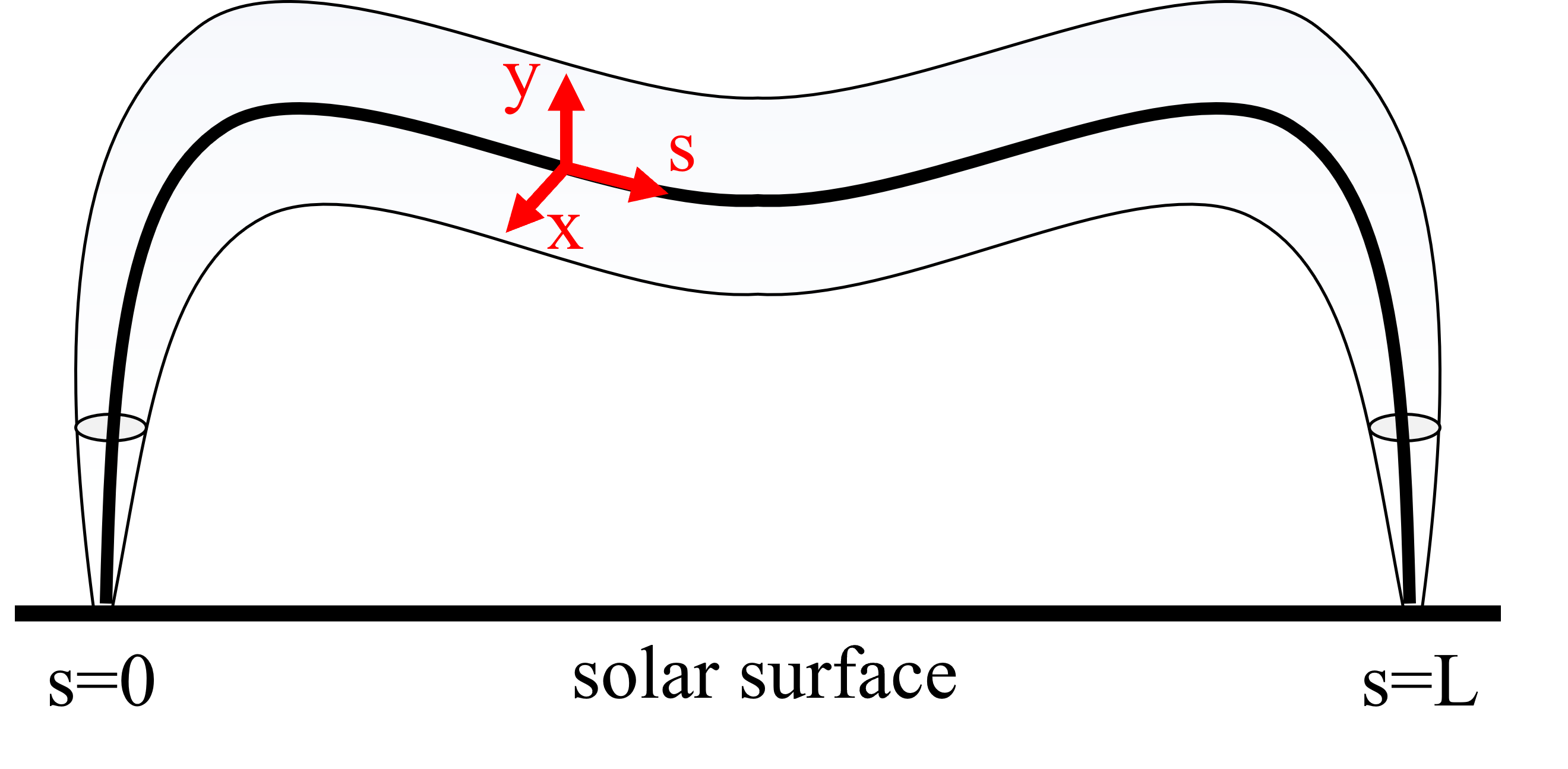}
  \caption{Schematic picture of the model.}
  \label{fig:loop}
\end{figure}

We numerically solve time-dependent MHD equations along a 1D magnetic field line with a dip, where `$s$' denotes the coordinate along the loop, and `$x$' and `$y$' represent directions perpendicular to the loop (Figure \ref{fig:loop}).
The smaller $s$ side is defined as the left side.
Phenomenological Alfv$\Acute{\mathrm{e}}$n wave turbulent dissipation is also considered \citep{shoda2018self}.
The basic equations are as follows:
\begin{gather}
    \frac{\partial}{\partial t}(\rho f_{\mathrm{ex}}) + \frac{\partial}{\partial s} (\rho v_s f_{\mathrm{ex}})=0, \label{mass} \\
    \frac{\partial}{\partial t} (\rho v_s f_{\mathrm{ex}}) + \frac{\partial}{\partial s} \left[\left(\rho v_{s}^{2} + p + \frac{\mathbf{B}_{\perp}^{2}}{8\pi}  \right) f_{\mathrm{ex}} \right] \nonumber \\
    = \left(p+\frac{\rho \mathbf{v}_{\perp}^{2}}{2} \right) \frac{d}{ds} f_{\mathrm{ex}} - \rho g f_{\mathrm{ex}}, \label{moment} \\
    \frac{\partial}{\partial t}(\rho \mathbf{v}_{\perp} f_{\mathrm{ex}}^{3/2}) + \frac{\partial}{\partial s}\left[\left(\rho v_s \mathbf{v}_{\perp} - \frac{B_s \mathbf{B}_{\perp}}{4\pi} \right)f_{\mathrm{ex}}^{3/2} \right] \nonumber \\
    = - \hat{\mathbf{\eta}}_{1}\cdot \rho \mathbf{v}_{\perp} f_{\mathrm{ex}}^{3/2} - \hat{\mathbf{\eta}}_{2} \cdot \sqrt{\frac{\rho}{4\pi}} \mathbf{B}_{\perp} f_{\mathrm{ex}}^{3/2}, \label{vperp} \\
    \frac{\partial}{\partial t} (\mathbf{B}_{\perp}\sqrt{f_{\mathrm{ex}}}) + \frac{\partial}{\partial s}[(\mathbf{B}_{\perp}v_{s} - B_s \mathbf{v}_{\perp})\sqrt{f_{\mathrm{ex}}}] \nonumber \\
    = - \hat{\mathbf{\eta}}_{1} \cdot \mathbf{B}_{\perp}\sqrt{f_{\mathrm{ex}}} - \hat{\mathbf{\eta}}_{2} \cdot \sqrt{4\pi\rho} \mathbf{v}_{\perp} \sqrt{f_{\mathrm{ex}}}, \label{Bperp} \\
    \frac{d}{ds}(f_{\mathrm{ex}} B_s) = 0, \label{divB} \\
    \frac{\partial}{\partial t}\left[\left(e + \frac{1}{2}\rho v^2 + \frac{\mathbf{B}^2}{8\pi} \right) f_{\mathrm{ex}} \right] \nonumber \\
    + \frac{\partial}{\partial s} \left[\left(e + p + \frac{1}{2}\rho v^2 + \frac{\mathbf{B}_{\perp}^{2}}{4\pi} \right) v_{s} f_{\mathrm{ex}} - B_{s} \frac{\mathbf{B}_{\perp}\cdot \mathbf{v}_{\perp}}{4 \pi}f_{\mathrm{ex}} \right] \nonumber \\
    = f_{\mathrm{ex}} (-\rho g v_{s} - Q_{\mathrm{R}} + Q_{\mathrm{C}} + Q_{\mathrm{L}}), \label{EnergyEq}
\end{gather}
where $\mathbf{v}$, $\mathbf{B}$, $\rho$, $p$, $e$, and $T$ are velocity, magnetic field, density, gas pressure, internal energy density, and temperature, respectively.
The transverse components of velocity and magnetic fields are given as $\mathbf{v}_{\perp} = v_{x}\mathbf{e}_{x}+v_{y}\mathbf{e}_{y} $, $\mathbf{B}_{\perp} = B_{x}\mathbf{e}_{x} + B_{y}\mathbf{e}_{y} $, where $\mathbf{e}_{x}$ and $\mathbf{e}_{y}$ represent unit vectors along $x$- and $y$-axis, respectively.
$p$ and $T$ are determined by following:
\begin{gather}
    e = \frac{p}{\gamma-1}, \label{energy} \\
    p = \frac{\rho k_{\mathrm{B}} T}{\mu m_{\mathrm{H}}}, \label{EoS}
\end{gather}
where $m_{\mathrm{H}}$ is the hydrogen mass, $k_{\mathrm{B}}$ is the Boltzmann constant, and $\gamma=5/3$ is the adiabatic index. $\mu$ represents the mean molecular weight.
For simplicity, we assume a fully ionized state consisting only of hydrogen and fix the value of $\mu=0.5$.

The gravitational acceleration along the loop $g(s)$ in the left half of the loop ($0\leq s \leq L/2$) is given by,
\begin{equation}\label{eq:gravity}
    g(s) =
    \begin{cases}
         g_{\mathrm{\odot}} \ (0\leq s <s_{1}), \\
         g_{\mathrm{\odot}} \cos({\frac{\pi}{2}\frac{s-s_{1}}{s_{2}-s_{1}}}) \ (s_{1}\leq s < s_{2}), \\
         g_{\mathrm{\odot}} \frac{\pi L_{\mathrm{dip}}}{2(L/2-s_{2})}  \sin{\pi \frac{s-s_{2}}{L/2-s_{2}}} \ (s_{2} \leq s<L/2),
    \end{cases}
\end{equation}
where $L=100$ Mm, $L_{\mathrm{dip}}=1$ Mm, $s_{1}=2$ Mm, $R_{\mathrm{s}}=8$ Mm, and $s_{2}=s_{1}+\pi R_{s}/ 2$ are the loop length, the depth, the location of the upper end of the coronal foot, the radius of the shoulder of a coronal loop, and the location of the shoulder. The gravitational acceleration is $g_{\mathrm{\odot}}=2.74\times 10^{4}\ \mathrm{cm\ s^{-2}}$.
The right half ($L/2< s \leq L$) is symmetric to the left half and is as follows,
\begin{equation}
    g(s) = -g(L-s).
\end{equation}
These formulae reflect the loop structure with the depth.

The variation in the cross-sectional area of the loop is given by the expansion factor $f_{\mathrm{ex}}(s)$ which is a function of $s$. $f_{\mathrm{ex}}(s)$ is assumed like below,
\begin{align}
    f_{\mathrm{ex}} = \left(\frac{1}{f_{\mathrm{ex,0}}^{10}} + \frac{1}{f_{\mathrm{ex,max}}^{10}} \right)^{-1/10},\label{eq:expansion factor2} \\
    f_{\mathrm{ex},0} = \min \left(\exp \left(\frac{\mu g_{\odot} m_{\mathrm{H}}}{2 k_{\mathrm{B}} T_{\mathrm{surf}}} \tilde{s} \right), f_{\mathrm{ex,max}} \right), \label{eq:expansion factor}
\end{align}
where $f_{\mathrm{ex,max}}=100$ represents the maximum value of the expansion factor over $s$.
$\tilde{s}$ represents the length along the loop from the closer footpoint given as,
\begin{gather}
    \tilde{s}(s) = \frac{L}{2} - \left|s- \frac{L}{2} \right|.
\end{gather}
$f_{\mathrm{ex}}$ depends on the scale height in the photosphere and chromosphere, but it becomes constant at a certain height above within the corona.
By using this expansion factor $f_{\mathrm{ex}}$, the magnetic field strength in Equation \eqref{divB} is determined as
\begin{gather}
    B_{s} = \frac{1}{f_{\mathrm{ex}}} B_{s,\mathrm{surf}},
\end{gather}
where $B_{s,\mathrm{surf}}=1000\ \mathrm{G}$. The plasma $\beta$ is constant in the photosphere and chromosphere.
The subscript `surf' denotes the value at the solar surface ($\tilde{s}=0$ Mm).
The minimum value of $B_{s}$ over $s$ is 10 G, consistent with observations of quiescent prominences \citep{leroy1984new, yamasaki2023magnetic}.

\begin{figure}[!]
  \epsscale{1.0}
  \plotone{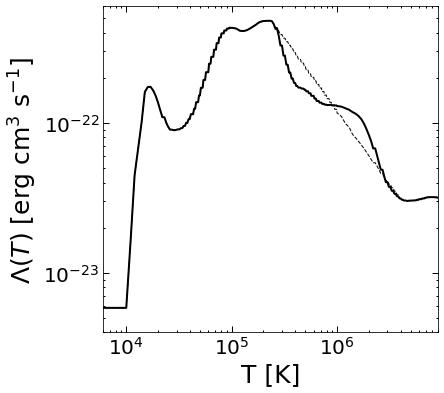}
  \caption{The optically thin radiative loss function $\Lambda (T)$. The dashed line represents a simplified used in Section \ref{sec:Field}.}
  \label{fig:radiativeloss}
\end{figure}

The source terms $\mathrm{Q_{R}}$, $\mathrm{Q_{C}}$ and $\mathrm{Q_{L}}$ in Equation (\ref{EnergyEq}) represent cooling (or heating) by the radiation, the thermal conduction and additional localized heating.
\cite{washinoue2022effect} demonstrated that chromospheric temperature plays a crucial role in corona modeling. To more realistically reproduce the solar atmosphere, the both optically thick and thin radiative cooling functions ($Q_{\mathrm{R,thick}}$ and $Q_{\mathrm{R,thin}}$) are applied to the cooling term $Q_{\mathrm{R}}$:
\begin{gather}
    Q_{\mathrm{R}} = Q_{\mathrm{R,thick}}(1-\xi_{\mathrm{t}}) 
                   + Q_{\mathrm{R,thin}} \xi_{\mathrm{t}}, \\
    \xi_{\mathrm{t}} = \min \left(1.0, \frac{p_{\mathrm{switch}}}{p} \right).
\end{gather}
The ratio between two terms is determined by $\xi_{\mathrm{t}}$ that is a measure of optical depth and is switched at  $p_{\mathrm{switch}}=1.0\ \mathrm{dyn\ cm^{-2}}$.
The optically thick radiative cooling $Q_{\mathrm{R,thick}}$ is approximated like below \citep{gudiksen2005ab},
\begin{gather}
    Q_{\mathrm{R,thick}} = 
    \frac{1}{\tau_{\mathrm{thick}}} (e-e_{\mathrm{ref}}), \\
    \tau_{\mathrm{thick}} = \tau_{\mathrm{thick,0}} \left(\frac{\rho}{\rho_{\mathrm{surf}}} \right)^{-\frac{1}{2}}, \\
    e_{\mathrm{ref}} = \frac{1}{\gamma-1}\frac{\rho k_{\mathrm{B}} T_{\mathrm{ref}}}{\mu m_{\mathrm{H}}}.
\end{gather}
$\tau_{\mathrm{thick}} $ is radiative cooling time. And $\rho_{\mathrm{surf}} = 10^{-7} \ \mathrm{g\ cm^{-3}}$, $\tau_{\mathrm{thick,0}}=0.1 \ \mathrm{s}$ and $e_{\mathrm{ref}}$ is internal energy at $T_{\mathrm{ref}}=4000\ \mathrm{K}$.
The optically thin radiative cooling $Q_{\mathrm{R,thin}}$ is divided into two components: the chromospheric and coronal terms, referred to as 
\begin{gather}
    Q_{\mathrm{R,thin}} = n_{\mathrm{H}} n_{\mathrm{e}} \Lambda(T)(1-\xi_{\mathrm{thin}}) + Q_{\mathrm{GJ}}(\rho,T) \xi_{\mathrm{thin}}, \\
    \xi_{\mathrm{thin}} = \max \left(0, \min \left(1, \frac{T_{\mathrm{TR}}-T}{\delta T} \right) \right),
\end{gather}
where $T_{\mathrm{TR}}=15000 \ \mathrm{K}$ is temperature at the transition region and $\delta T = 5000 \ \mathrm{K}$. Because of the fully ionized atmosphere, the number density of hydrogen $n_{\mathrm{H}}$ and electron $n_{\mathrm{e}}$ can be described as
\begin{gather}
    n_{\mathrm{H}} = n_{\mathrm{e}} = \frac{\rho}{m_{\mathrm{H}}}.
\end{gather}
$\Lambda (T)$ is the radiative loss function with photospheric abundances from Chianti Atomic Database ver. 7.0 \citep{dere1997chianti, landi2011chianti} (Figure \ref{fig:radiativeloss}) and $Q_{\mathrm{GJ}}(\rho, T)$ is the chromospheric radiative cooling function given by \citet{goodman2012radiating}.
For thermal conduction $Q_{\mathrm{C}}$, we adopt the Spitzer-H$\Ddot{\mathrm{a}}$rm conduction:
\begin{gather}
     Q_{\mathrm{C}} = \frac{1}{f_{\mathrm{ex}}}\frac{\partial}{\partial s}\left[(\kappa_{\mathrm{SH}}T^{5/2}) \frac{\partial T}{\partial s} f_{\mathrm{ex}}\right],
\end{gather}
where $\kappa_{\mathrm{SH}}=10^{-6} \ \mathrm{erg\ cm^{-1}\ s^{-1}\ K^{-7/2}}$ is the Spitzer conductivity \citep{spitzer1953transport}.
As shown later in Section \ref{sec:prominence formation}, after the loop reaches a quasi-steady state, an additional localized heating term $Q_{\mathrm{L}}$ is incorporated at footpoint(s) to cause the chromospheric evaporation. The formula is given as
\begin{equation}\label{eq:Localized heating}
    Q_{\mathrm{L}} (s,t) = Q_{\mathrm{local}}F(s)G(t),
\end{equation}
where $Q_{\mathrm{local}}$ is the footpoint heating rate at its maximum. $F(s)$ and $G(t)$ are functions of spatial and temporal distributions.
We assume that a single transient heating occurs at a single footpoint of the loop.
$F(s)$ and $G(t)$ are like below \citep{huang2021unified} (Type A),
\begin{gather}
    F(s) = \exp{\left( - \frac{(s-s_{\mathrm{peak}})^{2}}{\ell^{2}} \right)} \label{eq:Localized heating space AB} \\
    G(t) = \sum_{j=1,2} \exp{\left(- \frac{(t-t_{\mathrm{peak}})^{2}}{\tau_{j}^{2}} \right)} \mathrm{step}((-1)^{j}(t-t_{\mathrm{peak}})) \label{eq:Localized heating time AB},
\end{gather}
where $s_{\mathrm{peak}}$ and $t_{\mathrm{peak}}$ are where and when the heating term is at its maximum.
$\ell=0.15$ Mm is the heating width. $\tau_{\mathrm{1}}=100$ s and $\tau_{\mathrm{2}}=300$ s are the growing and decaying timescale.
$\mathrm{step}(\alpha)$ is a step function that is 0 when $\alpha<0$ and 1 when $\alpha>0$.
To compare with previous studies, three types of both-sided heating are also applied.
The first involves applying the same Type A heating to both footpoints of the loop (Type B).
The second is steady heating that decays exponentially with height given as \citep[e.g.,][]{xia2011formation} (Type D),
\begin{gather}
    F(s) =
    \left\{
    \begin{array}{ll}
    1 & \tilde{s} \leq s_{\mathrm{tr}}\\
    \exp{[-(\tilde{s}-s_{\mathrm{tr}}}) / \ell_{\mathrm{exp}}] \  & s_{\mathrm{tr}} < \tilde{s} \leq L/2
    \end{array}
    \right. \label{eq:Localized heating space C} \\
    G(t) = 
    \left\{
    \begin{array}{ll}
    (t-t_{\mathrm{start}})/\tau_{\mathrm{lin}} & t \leq t_{\mathrm{start}}+\tau_{\mathrm{lin}}\\
    1 & t > t_{\mathrm{start}}+ \tau_{\mathrm{lin}}
    \end{array}
    \right. \label{eq:Localized heating time C},
\end{gather}
where $s_{\mathrm{tr}}$ is the height of the transition region. $\ell_{\mathrm{exp}}$ is a heating scale height. $t=t_{\mathrm{start}}$ is the start time for calculating prominence formation.
This localized heating is set to increase linearly with the timescale $\tau_{\mathrm{lin}}$.
The third type adopts the spatial distribution described by Equation \eqref{eq:Localized heating space AB} and the temporal distribution described by Equation \eqref{eq:Localized heating time C} (Type C).

Dissipation via turbulent cascade is considered phenomenologically \citep{shoda2018self}. $\hat{\eta}_{1}$ and $\hat{\eta}_{2}$, coefficient tensors in Equations (\ref{vperp}) and (\ref{Bperp}), are implemented like below:
\begin{gather}
    \hat{\eta}_{1} = \frac{c_{d}}{4\lambda_{\mathrm{cor}}}
    \left(
    \begin{array}{cc}
    |\xi_{x}^{+}|+|\xi_{x}^{-}|&\	0\\
    0&\	|\xi_{y}^{+}|+|\xi_{y}^{-}|
    \end{array}     \right), \label{eta1} \\
    \hat{\eta}_{2} = \frac{c_{d}}{4\lambda_{\mathrm{cor}}}
    \left(
    \begin{array}{cc}
    |\xi_{x}^{+}|-|\xi_{x}^{-}|&\	0\\
    0&\	|\xi_{y}^{+}|-|\xi_{y}^{-}|
    \end{array}     \right), \label{eta2}
\end{gather}
where $\xi_{x}^{\pm}$ and $\xi_{y}^{\pm}$ are Elsasser variables \citep{elsasser1950hydromagnetic}:
\begin{gather}
    \mathbf{\xi^{\pm}} = \mathbf{v}_{\perp} \mp \frac{\mathbf{B}_{\perp}}{\sqrt{4\pi \rho}}.
\end{gather}
$\lambda_{\mathrm{cor}}$ is the correlation length perpendicular to the mean field. We assume it increases with the expansion of the flux tube:
\begin{equation}
    \lambda_{\mathrm{cor}} = \lambda_{\mathrm{cor,surf}} \sqrt{\frac{B_{s,\mathrm{surf}}}{B_{s}}},
\end{equation}
where $\lambda_{\mathrm{cor,surf}}=100\ \mathrm{km}$ is a typical length of inter granular lanes. We adopt $c_{d}=0.25$.

\subsection{Initial Condition}\label{sec:initial condition}
In the first step of our simulations, we reproduce a steady corona in thermal equilibrium. The initial conditions are as follows:
\begin{gather}
    T(s) = T_{\mathrm{surf}}, \label{eq:ICtemp} \\
    \rho (s) = \rho_{\mathrm{surf}} \exp{\left(-\frac{\min (\tilde{s}, s_{\mathrm{cor}})}{H} \right)}, \label{eq:ICdensity} \\
    p(s) = \frac{\rho(s) k_{\mathrm{B}} T_{\mathrm{surf}}}{\mu m_{\mathrm{H}}}, \label{eq:ICpressure}
\end{gather}
where $T_{\mathrm{surf}}=6000\ \mathrm{K}$, $\rho_{\mathrm{surf}} = 10^{-7} \ \mathrm{g\ cm^{-3}}$ and $\tilde{s}_{\mathrm{cor}}=7.5\times 10^{8} \ \mathrm{cm}$.
$H$ represents the gas pressure scale height given as,
\begin{equation}\label{scaleheight}
    H = \frac{k_{\mathrm{B}} T}{\mu g_{\odot }m_{\mathrm{H}}}.
\end{equation}
After the loop reaches a quasi-steady state, the resulting physical quantities are adopted as the initial conditions for calculating prominence formation.

\subsection{Boundary Condition}

The density and temperature at the boundary are fixed at $\rho_{\mathrm{surf}}$ and $T_{\mathrm{surf}}$.
To account for the effect of surface motion, transverse and longitudinal velocity perturbations with pink noise are injected:
\begin{gather}
    v_{i} \propto \sum_{N=0}^{N_{i}} \sin{(2\pi f_{N,i} t + \phi_{N,i} )}/ \sqrt{f_{N,i}} , \label{transboundary} \\
    f_{N,i} = f_{\min, i} + \frac{f_{\max,i}- f_{\min,i}}{N_{i}} N,\ \ (i=s,x,y) 
\end{gather}
where $N_{s}=10, N_{x,y}=20$. The frequency ranges are determined as $f_{\min,x}=f_{\min,y}=1.0\times 10^{-3} \ \mathrm{Hz}$, $f_{\max,x}=f_{\max,y}=1.0\times 10^{-2} \ \mathrm{Hz}$,
$f_{\min,s}=3.33 \ \mathrm{Hz}\times 10^{-3}$, and $f_{\max,s}= 1.0\times 10^{-2} \ \mathrm{Hz}$. Each $\phi$ is a different random number.
The root mean square (RMS) value of each velocity component is set to be $ v_{s,\mathrm{RMS}} = 0.9 \ \mathrm{km \ s^{-1}} $ and $ v_{i,\mathrm{RMS}} = 1.2 \ \mathrm{km \ s^{-1}} \ (i=x,y) $, which are comparable to the observed photospheric motions \citep[e.g.,][]{matsumoto2010temporal}.
Non-reflecting boundaries are set for both velocity and magnetic field.

\subsection{Numerical Scheme}

The numerical scheme employs the four-step Runge-Kutta method \citep{vogler2005simulations} and the fourth-order central finite difference method with artificial viscosity \citep{rempel2009radiative}. Thermal conduction is handled using the super-time stepping method \citep{meyer2012second, meyer2014stabilized}. We use a uniform grid size of $10 \ \mathrm{km}$.

\subsection{Simulation Cases}\label{Sec:simulation cases}
\begin{table*}[t]
\centering
  \begin{threeparttable}
    \caption{Parameters of the localized heating terms}
    \label{table:paramsurvey}
    \begin{tabular}{ccccc}
      \hline
      Type & both/one-sided & $F(s)$ & $G(t)$ & $Q_{\mathrm{local}} \ \mathrm{[erg \ cm^{-3} \ s^{-1}]}$ \\
      \hline \hline
      A & one  & Gaussian\tnote{1} & single \tnote{2} & 5.0, 16.0, 17.0, 30.0, 100.0  \\
      B & both & Gaussian & single & 2.0, 3.7, 3.8, 8.0  \\
      C & both & Gaussian & steady \tnote{3} & $5\times 10^{-4}, \ 5\times 10^{-3}, \ 1\times 10^{-2}, \ 5\times 10^{-2}$  \\
      D & both & exponential\tnote{4} & steady & $5\times10^{-4}, \ 2\times10^{-3}, \ 3\times10^{-3}, \ 5\times10^{-3}$ \\
      \hline
    \end{tabular}
    \begin{tablenotes}
    \item \makebox[0pt][c]{\hspace{-\tabcolsep} \hspace{8.2cm} 1. Eq.\eqref{eq:Localized heating space AB}}
          \hspace{6.0cm} \makebox[0pt][c]{2. Eq.\eqref{eq:Localized heating time AB}}%
          \hspace{2cm} \makebox[0pt][c]{3. Eq.\eqref{eq:Localized heating time C}}%
          \hspace{2cm} \makebox[0pt][c]{4. Eq.\eqref{eq:Localized heating space C}}
  \end{tablenotes}
  \end{threeparttable}
\end{table*}
\vspace{0.2cm} 
To investigate the mechanism of condensation with a single heating event and to compare it with previous studies with steady heating, we conduct simulations as detailed in Table \ref{table:paramsurvey}.
For Type A and B, a temporally single, spatially Gaussian heating is applied at the left side and both sides of the loop.
In Type A, we consider five cases with $Q_{\mathrm{local}}= 5.0, \ 16.0, \ 17.0, \ 30.0, \ 100.0 \ \mathrm{erg \ cm^{-3} \ s^{-1}}$ (Case A1-5), and in Type B, we consider four cases with $Q_{\mathrm{local}}= 2.0, \ 3.7, \ 3.8, \ 8.0 \ \mathrm{erg \ cm^{-3} \ s^{-1}}$ (Case B1-4).
For steady heating, Type C uses a spatially Gaussian heating function similar to the previous cases, with heating rates $Q_{\mathrm{local}} = 5\times 10^{-4}, \ 5\times10^{-3}, \ 1\times10^{-2}, \ 5\times10^{-2} \ \mathrm{erg\ cm^{-3}\ s^{-1}}$ (Case C1-4).
While Type D employs an exponentially decaying heating function, with rates $Q_{\mathrm{local}}=5\times 10^{-4}, \ 2\times 10^{-3}, \ 3\times 10^{-3}, \ 5\times10^{-3}, \ \mathrm{erg\ cm^{-3}\ s^{-1}}$

\begin{figure}[!]
  \epsscale{1.2}
  \plotone{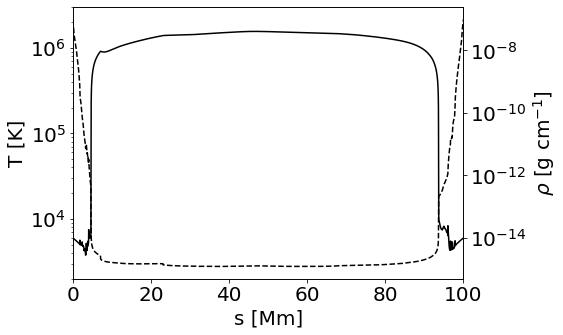}
  \caption{The temperature and density profiles at the quasi-steady state ($t=35,000$ s) are plotted by solid and dashed lines, respectively. }
  \label{fig:IC_prominence}
\end{figure}

\section{Results}\label{sec:results}

After the loop reaches a quasi-steady state, the resulting physical quantities are used as the initial conditions for the prominence formation simulations.
Section \ref{sec:Coronal heating problem} presents a brief overview of the results for the corona reproduction. The results for prominence formation in the standard case are discussed in Section \ref{sec:prominence formation}.
A parameter survey varying the magnitude of the localized heating is detailed in Section \ref{sec:parameter survey in the amplitude}.

\subsection{Reproduction of Corona}\label{sec:Coronal heating problem}

\begin{figure}[!]
  \epsscale{1.1}
  \plotone{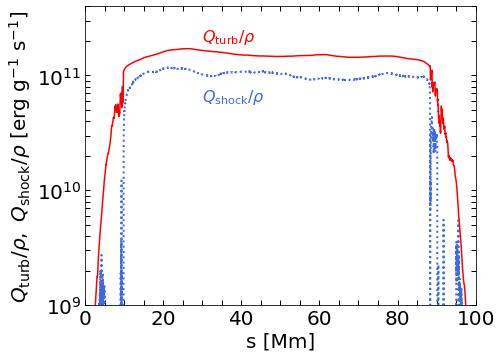}
  \caption{The profile of the heating rates due to shock waves $Q_{\mathrm{shock}}$ (blue dotted line) and turbulent cascades $Q_{\mathrm{turb}}$ (red solid line) are plotted. These rates are averaged over 65,000 s after the loop reaches a quasi-steady state.}
  \label{fig:coronaheating_HR}
\end{figure}

Figure \ref{fig:IC_prominence} illustrates a snapshot of the density and temperature profiles at the quasi-steady state, $t=t_{\mathrm{steady}}=35,000$ s.
At this moment, the density and temperature reach approximately $1.6\times 10^{-15} \ \mathrm{g \ cm^{-3}}$ and $1.3 \ \mathrm{MK}$, which are typical values for the corona.
In Figure \ref{fig:coronaheating_HR}, the red and blue lines represent the heating rate per unit mass along the loop due to shock waves and turbulence cascade, averaged over 50,000 s starting from $t=t_{\mathrm{steady}}$.
These rates are derived as follows.
The energy equation is \citep{shoda2018self}
\begin{gather}
    \frac{\partial e}{\partial t} = - v_{s} \frac{\partial e}{\partial s} - \frac{e+p}{f_{\mathrm{ex}}}\frac{\partial}{\partial s} (v_{s}f_{\mathrm{ex}}) \nonumber \\
    + Q_{\mathrm{C}} - Q_{\mathrm{R}} + Q_{\mathrm{shock}} + Q_{\mathrm{turb}}.
\end{gather}
\begin{figure*}[!]
  \epsscale{1.0}
  \plotone{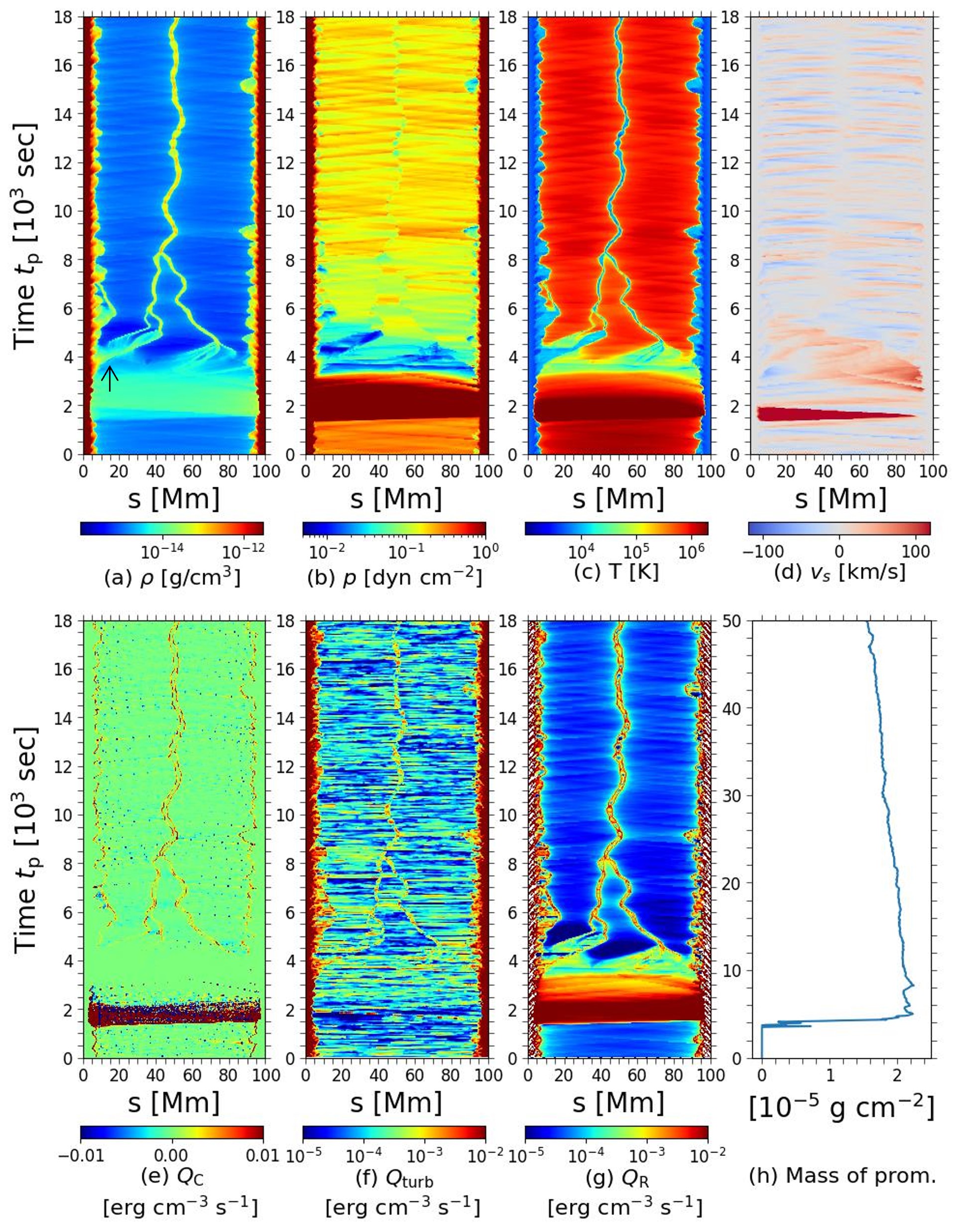}
  \caption{(a-g) Space-time diagrams for density, pressure, temperature, velocity along the loop $v_{s}$, thermal conduction $Q_{\mathrm{C}}$, turbulent heating $Q_{\mathrm{turb}}$ and cooling rate $Q_{\mathrm{R}}$ in Case A4 are shown from $t_{\mathrm{p}}=0$ to $1.8\times 10^{4}$ s. The horizontal and vertical axes represent the distance from the left footpoint of the loop and time with units of Mm and $10^{3}$ s. The initial conditions are the same as those in Figure \ref{fig:IC_prominence}. (h) The time evolution of the total mass of the prominence from $t_{\mathrm{p}}=0$ to $5.0\times 10^{4}$ s is shown. The horizontal axis represents mass per unit cross-sectional area. The black arrow in panel (a) indicates the tall spicule extending from the left transition region.}
  \label{fig:prominence formation}
\end{figure*}
$Q_{\mathrm{shock}}$ and $Q_{\mathrm{turb}}$ represent the heating rates due to shock waves and turbulence cascade, respectively.
$Q_{\mathrm{turb}}$ is expressed analytically as \citep{cranmer2007self, verdini2007Alfven}
\begin{equation}
\label{Qturb}
    Q_{\mathrm{turb}} = \frac{c_{d}}{4\lambda_{\mathrm{cor}}} \rho \sum_{i=x,y} (|\xi_{i}^{+}|{\xi_{i}^{-}}^{2} + |\xi_{i}^{-}|{\xi_{i}^{+}}^{2}).
\end{equation}
On the other hand, $Q_{\mathrm{shock}}$ is calculated as a time-averaged quantity:
\begin{gather}
    \overline{Q_{\mathrm{shock}}} = \overline{ \frac{\partial e}{\partial t} + v_{s} \frac{\partial e}{\partial s}
    + \frac{e+p}{f_{\mathrm{ex}}} \frac{\partial}{\partial s} (v_{s} f_{\mathrm{ex}}) } \\
    \overline{ - Q_{\mathrm{C}} + Q_{\mathrm{R}} - Q_{\mathrm{turb}} }. \nonumber
    \label{eq:Qshockave}
\end{gather}
It is found that turbulent heating, $Q_{\mathrm{turb}}/\rho$, is approximately twice as large as shock heating, $Q_{\mathrm{shock}}/\rho$, at the center of the loop, reaching around $10^{11} \ \mathrm{erg \ g^{-1} \ s^{-1}}$, as shown in Figure \ref{fig:coronaheating_HR}. The corresponding heat flux can be calculated as
\begin{gather}
    Q_{\mathrm{turb}} L_{\mathrm{corona}}\sim 10^{6} \ \mathrm{erg \ cm^{-2} \ s^{-1}}.
\end{gather}
This value corresponds to the coronal energy loss flux in the quiet Sun \citep{withbroe1977mass}, indicating that our numerical setting reproduces the coronal loop in the quiet regions.

The maximum vertical velocity amplitude along the loop is 29 $\mathrm{km/s}$, which reduces to approximately 20 $\mathrm{km/s}$ when considering a single component.
This result is consistent with the observed non-thermal velocities of around 20 $\mathrm{km/s}$ \citep[e.g.][]{hara1999microscopic}.

\subsection{Prominence Formation}\label{sec:prominence formation}

After the loop reaches a quasi-steady state at $t=t_{\mathrm{steady}}$, we introduce the localized heating $Q_{\mathrm{L}}$ at the left side of the coronal loop (Type A). 
A new time axis $t_{\mathrm{p}}$ is defined as $t_{\mathrm{p}}=t-t_{\mathrm{steady}}$. 
The temporal peak of the localized heating occurs at $t_{p}=t_{\mathrm{peak}}=1,500$ s, while the spatial peak, $s=s_{\mathrm{peak}}=8.5 \ \mathrm{Mm}$, is located approximately 1 Mm above the transition region, which is positioned at $s\sim7.5$ Mm.

\subsubsection{Overview of the standard case}\label{sec:overview of prominence formation}

We present the results of Case A4 in Figure \ref{fig:prominence formation}(a-g) as a representative example.
The peak amplitude of the localized heating, $Q_{\mathrm{local}}$, in Equation \eqref{eq:Localized heating} is set to $30.0 \ \mathrm{erg \ cm^{-3}\ s^{-1}}$.
Assuming a heating cross-sectional area of $0.15 \ \mathrm{Mm}\times 0.15 \ \mathrm{Mm}$, based on the heating distribution in \cite{robinson2022incoherent}, the total heating amount is $\sim 10^{25} \ \mathrm{erg}$.

Following the localized heating, the coronal temperature increases (Figure \ref{fig:prominence formation}(c)).
The density also rises due to chromospheric evaporation (Figure \ref{fig:prominence formation}(a)).
The evaporation flow is visible in the space-time diagram of $v_{\mathrm{s}}$ in Figure \ref{fig:prominence formation}(d), where an upflow occurs on the left side of the coronal loop.
Subsequently, the entire corona cools down to the transition region temperatures ($\sim 10^{5} \ \mathrm{K}$), resulting in local condensation between $3,300< t_{\mathrm{p}}<4,600 \ \mathrm{s}$ (Figure \ref{fig:prominence formation}(c)).

A tall spicule extends from the left transition region at $t_{\mathrm{p}}\sim 3,700$ s (indicated by the black arrow in Figure \ref{fig:prominence formation}(a)).
As the entire loop experiences runaway cooling between $3,300 < t_{p} < 3,700 \ \mathrm{s}$ (Figure \ref{fig:prominence formation}(c)), the accompanying drop in coronal pressure causes the transition region to rise (Figure \ref{fig:prominence formation}(b)).
Although not shown in the figure, it is confirmed that slow shocks generated by the boundary motion interact with and repeatedly lift the transition region when examining the footpoint of this spicule.
As a result, the spicule gradually ascends while oscillating between upward and downward motions.
In previous studies, such phenomenon was attributed to a rebound shock train associated with a single shock wave \citep{hollweg1982origin, sterling1988rebound}.
However, the primary causes in our study are intermittent shock waves generated by photospheric motion and the associated rebound shocks.
When the spicule reaches the shoulder of the loop, it falls into the dip and coalesces into the prominence.

The prominence exhibits oscillatory motion along the loop with a period of $2,000 - 3,500$ s.
This motion is interpreted as a manifestation of a pendulum-like oscillation \cite[e.g.,][]{jing2003periodic,jing2006periodic,luna2012formation,zhang2012observations}, with an estimated period of $\sim3,300$ s, considering the effects of gravity and the pressure gradient.

Figure \ref{fig:prominence formation}(h) shows the time evolution of prominence mass from $t_{\mathrm{p}}=0$ to $5.0\times10^{4}$ s.
The prominence mass is defined as the total mass of plasma with temperature $T<20,000$ K between the left and right transition regions.
Our simulation reveals a descending trend in prominence mass.
The average mass loss rate is approximately $\sim 1.3\times10^{-10} \ \mathrm{g \ cm^{-2} \ s^{-1}}$.
To ensure that this decrease is not caused by numerical diffusion, we confirmed that mass conservation holds within the region including the prominence.
This result contrasts with previous studies, which reported a persistent increase in mass even after the localized heating ceased \citep[e.g.,][]{xia2011formation}.
The cause of this discrepancy will be discussed in Section \ref{Relationship with previous}.
\begin{figure}[!]
  \epsscale{1.1}
  \plotone{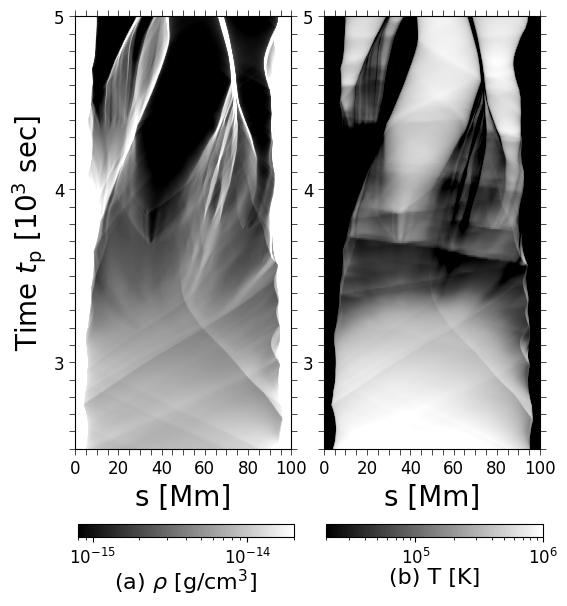}
  \caption{Space-time diagrams of density and temperature from $t_{\mathrm{p}}=2,500$ to $5,000$ s are shown. The horizontal axis represents the distance from the left footpoint of the loop. To enhance the visibility of the condensation process, the color bar ranges have been adjusted compared to those in Figure \ref{fig:prominence formation}.}
  \label{fig:zoomcondensation}
\end{figure}
\begin{figure*}[!]
  \epsscale{1.18}
  \plotone{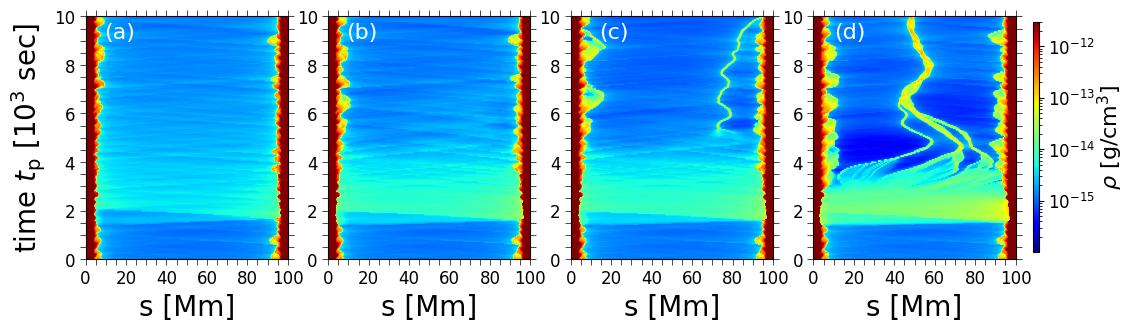}
  \caption{Space-time diagrams of density in Case (a) A1, (b) A2, (c) A3, and (d) A5. The initial condition at $t_{\mathrm{p}}=0$ s is same with the condition in Figure \ref{fig:prominence formation}.}
  \label{fig:TBn_compare_ro}
\end{figure*}

\subsubsection{Initial Stage of Condensation}\label{sec:InitialStageofCondensation}

Figure \ref{fig:zoomcondensation} represents space-time diagrams of density and temperature from $t_{\mathrm{p}}=2,500$ to $5,000$ s.
Frequent shock wave propagation is observed in the corona.
We propose that Alfv$\Acute{\mathrm{e}}$n waves generated by the boundary motion undergo mode conversion into longitudinal waves, which steepen into shock waves \citep[e.g.,][]{hollweg1982possible}.
The dense plasma in the downstream regions of these shock waves become even denser as the shock waves pass through each other.
This further increase in density in the downstream regions, in addition to the mass supply by chromospheric evaporation, enhances radiative cooling, leading to localized condensation.
A similar process was observed by \cite{antolin2010coronal}.

\begin{figure}[!]
  \epsscale{1.2}
  \plotone{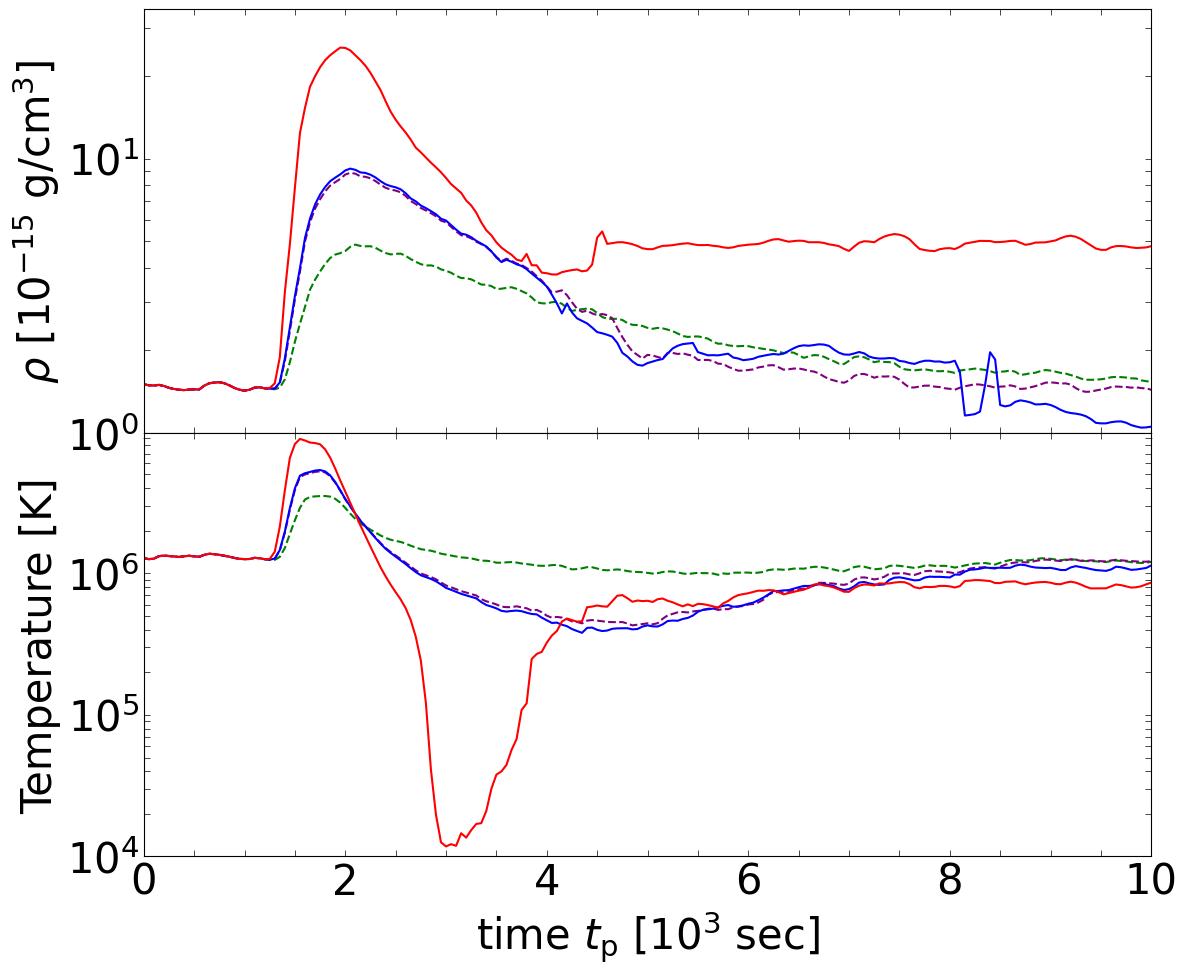}
  \caption{The time evolution of coronal averaged density (top) and temperature (bottom) for each case in the parameter survey is shown. Red and blue solid lines represent Case A5 and A3, which exhibit condensations, while purple and green dashed lines represent Case A2 and A1, which do not show any condensations.}
  \label{fig:compare_ro_Te}
\end{figure}

\subsection{Parameter Survey on Magnitude of Localized Heating}\label{sec:parameter survey in the amplitude}

To investigate the condition for condensation, we perform a parameter survey varying the localized heating rate, $Q_{\mathrm{local}}$, as described in Section \ref{Sec:simulation cases} (Case A1-3, 5).
We found that condensation occurs when $Q_{\mathrm{local}}=17.0 \ \mathrm{erg \ cm^{-3} \ s^{-1}}$ (Case A3) or higher (Figure \ref{fig:TBn_compare_ro}).

Figure \ref{fig:compare_ro_Te} shows the time variation of the averaged density and temperature in the corona.
In each case, the temperature reaches its maximum around $t_{\mathrm{p}}= 1,800$ s, with the density peaking slightly later.
In Case A5, both density and temperature decrease rapidly from $t_{\mathrm{p}}\sim 2,500$ s, and condensation occurs at $t_{\mathrm{p}}\sim 2,900$ s.
In contrast, in Case A1, both density and temperature decrease monotonically to levels observed before the localized heating.
Cases A2 and A3 exhibit similar behavior, though condensation occurs only in Case A3. The temperature profile shifts from decreasing to increasing around $t_{\mathrm{p}}\sim 3,950$ s in both cases.
The conditions for condensation are discussed in more detail in Section \ref{sec:Field}.

\section{Discussion} \label{sec:Discussion}  

\subsection{Condition for Condensation}\label{sec:Field}

One of the key factors influencing condensation is the interval time $\tau_{\mathrm{int}}$ between intermittent localized heating events. 
Previous studies have examined this condition for steady \citep[e.g.,][]{antiochos1999dynamic,xia2011formation}, periodic \citep[e.g.,][]{karpen2008condensation, johnston2019effects}, episodic \citep[e.g.,][]{antolin2010coronal}, or random \citep[e.g.,][]{kucera2024modeling} localized heatings.
It has been reported that $\tau_{\mathrm{int}}$ must be shorter than radiative cooling time $\tau_{\mathrm{cool}}=k_{\mathrm{B}}T/[(\gamma-1)n\Lambda (T)]$ to maintain the coronal density during the cooling process \citep{karpen2008condensation,johnston2019effects}, which is expressed as
\begin{equation}\label{eq:tau_int_tau_R}
    \tau_{\mathrm{int}} < \tau_{\mathrm{cool}}.
\end{equation}
We will follow convention and refer to such localized heating as quasi-steady heating.
Quasi-steady heating is regarded as the accumulation of multiple heating events occurring across different field lines.
However, it is reasonable to consider condensation triggered by a single heating event along a single field line as an elemental unit.
In this scenario, the interval time $\tau_{\mathrm{int}}$ can be considered extremely large, exceeding the cooling time $\tau_{\mathrm{cool}}$, i.e., $\tau_{\mathrm{int}}\gg\tau_{\mathrm{cool}}$.
Several numerical studies have investigated condensation under this condition \citep{reep2020electron,kohutova2020self,huang2021unified}, yet the underlying mechanism remains poorly understood.

\begin{figure}[!]
  \epsscale{1.2}
  \plotone{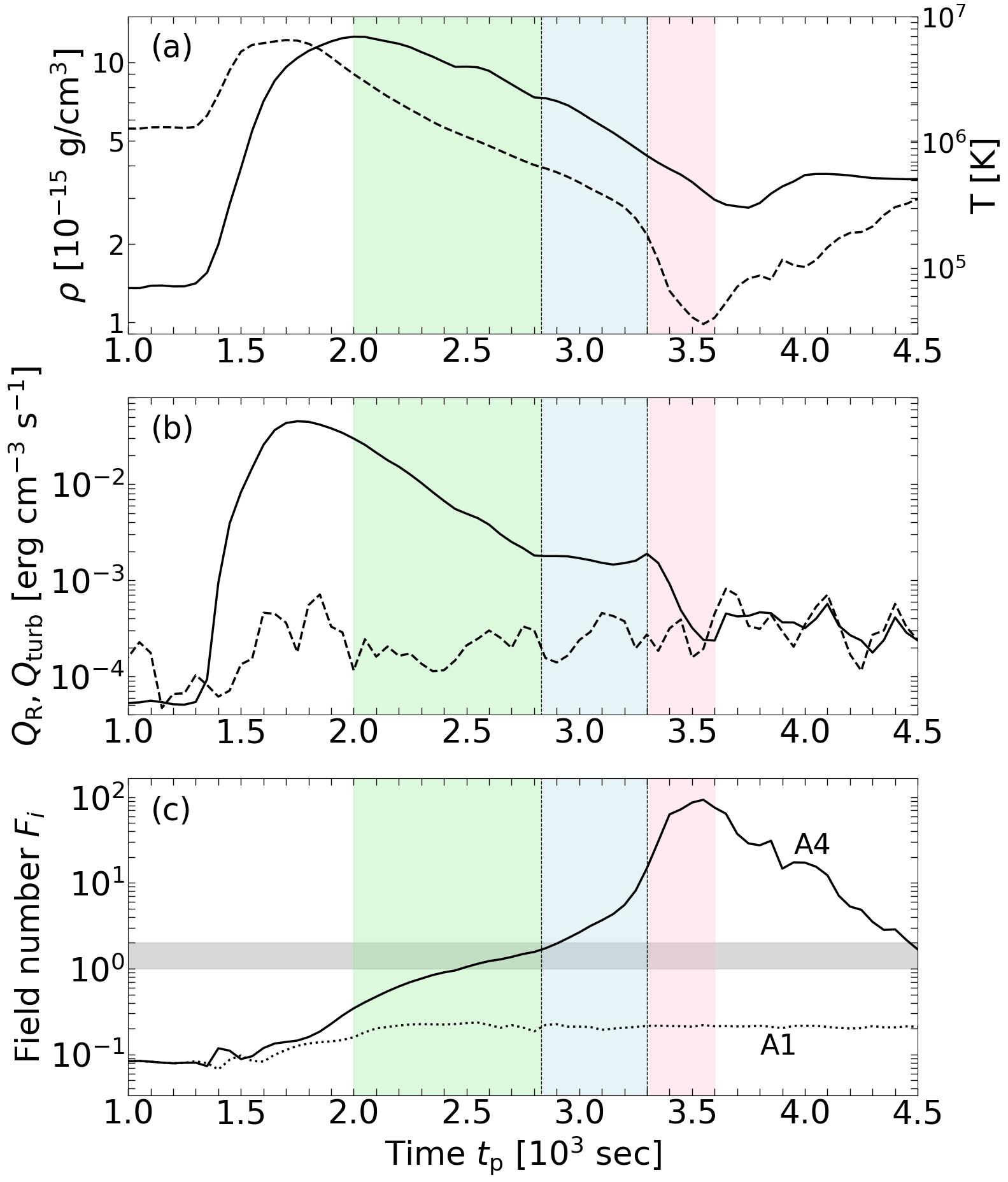}
  \caption{The time variations of (a) density (solid line) and temperature (dashed line), (b) radiative cooling rate $Q_{\mathrm{R}}$ (solid line) and turbulent heating $Q_{\mathrm{turb}}$ (dashed line) in Case A4, and (c) the Field number $F_{\mathrm{i}}$ in Case A4 (solid line) and A1 (dotted line) are shown. The time from the maximum density to the onset of condensation in Case A4 is divided into three phases: $t_{\mathrm{p}}\approx 2,000-2,800$ s (green), $t_{\mathrm{p}}\approx 2,800-3,300$ s (blue), and $t_{\mathrm{p}}\approx 3,300-3,600$ s (red).}
  \label{fig:compare_roTe_QR_LF}
\end{figure}

Figure \ref{fig:compare_roTe_QR_LF} plots the time evolution of (a) density and temperature, (b) radiative cooling rate $Q_{\mathrm{R}}$ and turbulent heating rate $Q_{\mathrm{turb}}$, and (c) the Field number $F_{\mathrm{i}}$, averaged over the corona in Case A4.
Here, we define the following variable as the Field number $F_{\mathrm{i}}$:
\begin{equation}\label{Field number}
    F_{\mathrm{i}} = \frac{L}{2}\sqrt{\frac{n^{2}\Lambda(T)}{4\pi^{2} \kappa_{0}T^{7/2}}}.
\end{equation}
$F_{\mathrm{i}}$ is a dimensionless number related to the Field length $\lambda_{\mathrm{F}}$ \citep{field1965thermal}, expressed as
\begin{equation}
    \lambda_{\mathrm{F}} = 2\pi \sqrt{\frac{\kappa_{0} T^{7/2}}{n^{2}\Lambda (T)-Q_{\mathrm{heat}}}},
\end{equation}
representing the ratio of the efficiency of the effective cooling to that of thermal conduction in a coronal loop.
$Q_{\mathrm{heat}}$ represents the total heating rate.
Since the turbulent heating rate $Q_{\mathrm{turb}}$ during condensation is less than 1/5 of the cooling rate $Q_{\mathrm{R}}$ for most of the time in our simulations (see Figure \ref{fig:compare_roTe_QR_LF}(b)), $Q_{\mathrm{turb}}$ is excluded from Equation \eqref{Field number}.
When $F_{\mathrm{i}}>1$, the efficiency of thermal conduction along the loop becomes lower than the cooling efficiency.

The cooling profile exhibits a decreasing trend during $t_{\mathrm{p}}=2,000-2,800$ s, shifts to an increasing trend, and then decreases again starting from $t_{\mathrm{p}}=3,300$ s (Figure \ref{fig:compare_roTe_QR_LF}(b)).
To understand this behavior, we divide the time from the maximum density to the onset of condensation into three phases, based on the observed variations in the cooling rate.

Initially, from $t_{\mathrm{p}}=2,000$ to $2,800$ s (green), as the temperature and pressure scale height decrease, the density stratification changes, causing the coronal plasma to drain into the chromosphere ($d\rho/ dt<0$).
This leads to a reduction in the radiative cooling rate $Q_{\mathrm{R}}=n_{e}^{2}\Lambda(T)$ ($dQ_{\mathrm{R}}/dt<0$), despite the radiative loss function $\Lambda(T)$ increasing as the temperature decreases.
At $t_{\mathrm{p}}\sim 2,800$ s (blue), the cooling rate begins to rise (i.e., $dQ_{\mathrm{R}}/dt \gtrsim 0$).
This turning point coincides with when the Field number $F_{\mathrm{i}}$ reaches approximately 1.6.
The slight exceedance of $F_{\mathrm{i}}$ over 1 when condensation occurs in Case A4 is attributed to the fact that the effect of heating is not included in Equation \eqref{Field number}.
Additionally, this is due to averaging over the entire corona, which tends to obscure localized cooling phenomena.
Subsequently, both temperature and density decrease rapidly.
When the temperature drops to $T=0.15$ MK at around $t_{\mathrm{p}}=3,000$ s, the slope of the radiative loss function $\Lambda (T)$ shifts from negative to positive (see Figure \ref{fig:radiativeloss}), causing $dQ_{\mathrm{R}}/dt$ to turn negative again.
This behavior is also observed in other cases where condensation occurs.
Meanwhile, in Case A1 where no condensation is observed, the Field number $F_{\mathrm{i}}$ does not exceed 0.3 (Figure \ref{fig:compare_roTe_QR_LF}(c)).
\begin{figure*}[!]
  \epsscale{1.0}
  \plotone{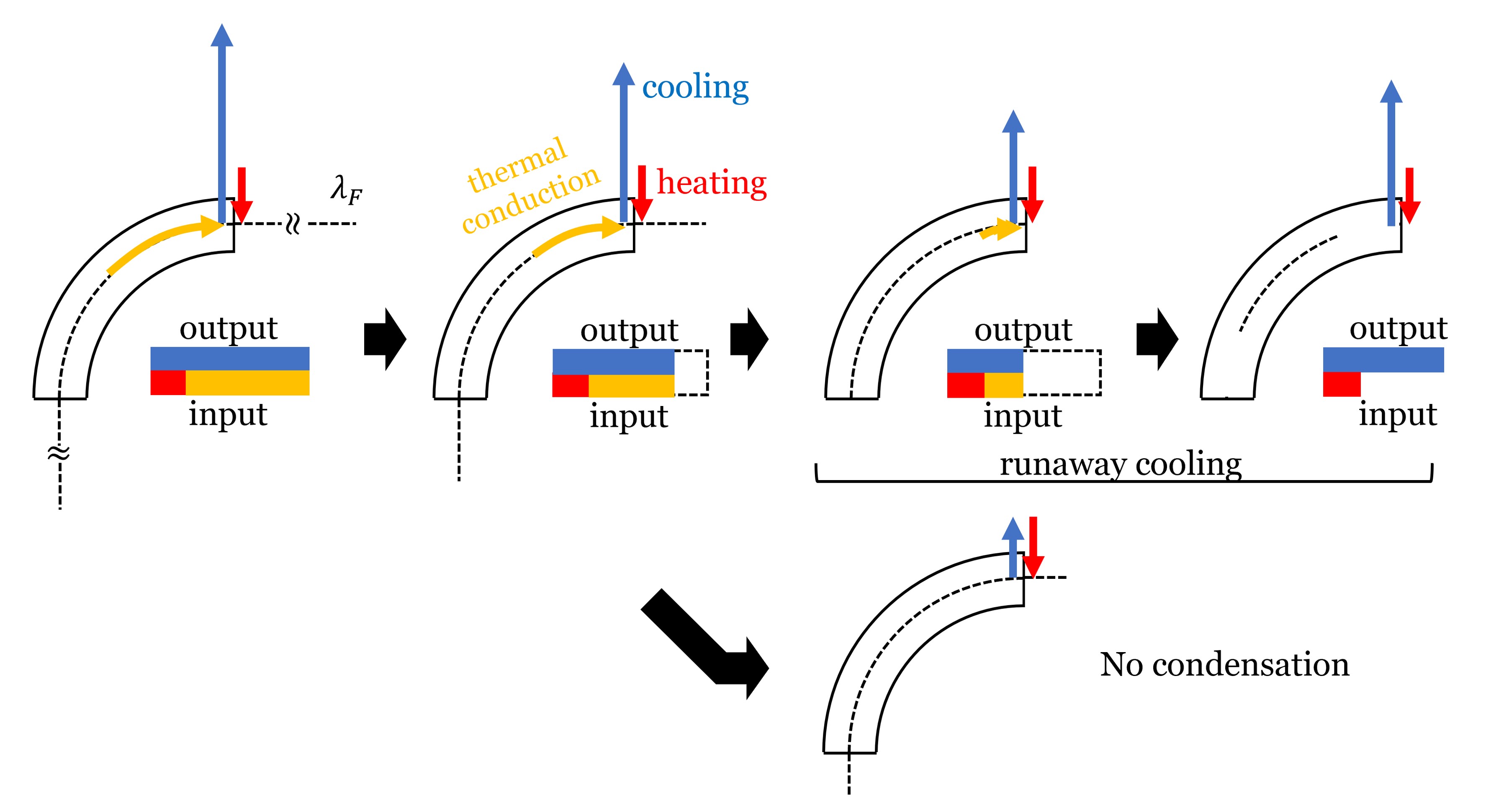}
  \caption{The schematic picture of two cases: (upper) condensation occurring due to a single heating event (refer to $t_{p}=2,000$ to $3,000$ s in Case A4) and (lower) no condensation occurring. Solid lines represent half of the loop, while dashed lines denote the Field length. The energy relationships between heating (red), cooling (blue), and thermal conduction (yellow) at the upper corona are indicated by arrows. Their lengths indicate the magnitude of energy input or output. The first panel depicts the situation after the localized heating.}
  \label{fig:final_image}
\end{figure*}
In summary, condensation in our simulations is consistently accompanied by the Field number exceeding unity,
\begin{equation}
\label{eq:new_condition_for_Field}
    F_{\mathrm{i}} \gtrsim 1.0.
\end{equation}
In this situation, thermal conduction is unable to stabilize perturbations in the system, resulting in runaway cooling.
This result is consistent with the previous studies, which showed condensation caused by an increase in number density $n$ due to steady localized heating \citep{xia2011formation} and by an increase in loop length $L$ through magnetic reconnection \citep{kaneko2017reconnection}, both of which lead to a larger Field number $F_{\mathrm{i}}$.

The schematic pictures of the energy balance between heating, cooling, and thermal conduction in the upper corona, for cases with and without condensation, are shown in Figure \ref{fig:final_image}.
Initially, the Field number $F_{\mathrm{i}}$ is less than unity.
Due to evaporation, cooling (blue arrow) significantly exceeds heating (red arrow), and excess cooling is balanced by thermal conduction (yellow arrow).
As the temperature decreases, $F_{\mathrm{i}}$ gradually increases.
This balance holds roughly until $F_{\mathrm{i}}$ approaches unity.
However, as the temperature continues to drop, thermal conduction becomes ineffective, and the excess cooling can no longer be compensated, leading to runaway cooling.
If the cooling rate decreases to match the heating rate under conditions where $F_{\mathrm{i}}<1.0$, the temperature drop halts, and the system returns to its pre-localized heating state.
Here, it is assumed that the heating rate remains largely unchanged.
In the non-equilibrium model \citep[e.g.,][]{antiochos1999dynamic}, coronal density continues to rise due to quasi-steady localized heating until the loop reaches a new equilibrium. If this new state satisfies $F_{\mathrm{i}}\gtrsim 1.0$, condensation occurs.

The radiative loss function in Figure \ref{fig:radiativeloss} exhibits significant changes in slope around temperatures of $T=0.4$ and $1.5$ MK, in addition to $T=0.15$ MK.
To ensure that these variations did not influence the results, an additional calculation was performed using a simplified loss function with a constant slope between $T=0.15$ and $4.0$ MK (represented by the dashed line in Figure \ref{fig:radiativeloss}).

\subsection{Required Amount of Localized Heating}\label{The effect of background heating}
The condition for thermal non-equilibrium has been studied by numerous researchers. 
An important factor is the ratio of localized heating to background heating heating \citep{antolin2010coronal, johnston2019effects,klimchuk2019role}.
\cite{klimchuk2019role} analytically derived the condition for thermal non-equilibrium as follows:
\begin{equation}
  \frac{Q_{\lambda_{\mathrm{H}}}}{Q_{\mathrm{min}}} > 1+\frac{c_{1}}{\Gamma_{\lambda_{\mathrm{H}}}}.\label{eq:KL2019}
\end{equation}
$Q_{\mathrm{min}}$ and $Q_{\lambda_{\mathrm{H}}}$ represent the volumetric heating rates at the loop apex and at one heating scale length $\lambda_{\mathrm{H}}$ above the transition region, respectively.
$c_{1}\equiv R_{\mathrm{TR}}/R_{\mathrm{cor}}$ denotes the ratio of the radiative losses per unit area $R$ integrated along the transition region to those in the corona sections.
$\Gamma_{\lambda_\mathrm{H}}\equiv A_{\lambda_{H}}/ A_{\mathrm{TR}}$ is the ratio of the cross-sectional areas $A$ of the transition region and at one heating scale length above the transition region.
Note that this condition assumes exponential decay of steady heating from the transition region applied to both footpoints.

If we calculate the required ratio of the localized heating rate to the background heating rate using the formula \eqref{eq:KL2019}, it ranges from 2.0 to 4.5.
The value of $c_{1}$ varies from 1.0 to 3.5, where $R_{\mathrm{TR}}$ and $R_{\mathrm{cor}}$ are evaluated by defining the corona and transition region as regions with $T>4\times10^{5}$ K and $1.5\times10^{4} \ \mathrm{K}<T<4\times10^{5} \ \mathrm{K}$, respectively.
$\Gamma_{\lambda_{\mathrm{H}}}\equiv A_{\lambda_{\mathrm{H}}}/A_{\mathrm{TR}}$ is calculated from the ratio of the expansion factors at $s=s_{\mathrm{peak}}$ and at the transition region.
In contrast, our parameter survey indicates that the actual ratio of the localized heating rate to the background heating rate required for condensation to occur was
\begin{equation}
    \frac{ Q_{\lambda_{\mathrm{H}}} }{ Q_{\mathrm{min}} }
    \sim \frac{Q_{\mathrm{local}}/e}{\bar{Q}_{\mathrm{bg}}} \sim 2 \times 10^{4}.
\end{equation}
Here, we extended the interpretation of $Q_{\mathrm{min}}$ as follows:
\begin{equation}
    \bar{Q}_{\mathrm{bg}} = \frac{\int_{s_{\mathrm{-},\mathrm{cor}}}^{s_{\mathrm{+},\mathrm{cor}}} ds \int_{t_{1}}^{t_{2}} dt \ Q_{\mathrm{bg}}(s,t)}{(s_{\mathrm{+,cor}}-s_{\mathrm{-,cor}})(t_{2}-t_{1})}.
\end{equation}
The integral spatial range is defined as $s_{\pm,\mathrm{cor}}=L/2 \pm \Delta s$ and $s_{\pm,\mathrm{local}}=s_{\mathrm{peak}} \pm \Delta s$, with $\Delta s=1 \ \mathrm{Mm}$.
The integral temporal range is defined as $t_{1}=t_{\mathrm{peak}}-\tau_{\mathrm{1}}$,
and $t_{2}=t_{\mathrm{peak}}+\tau_{\mathrm{2}}$.
The constants related to time ($t_{\mathrm{peak}}, \ \tau_{1}, \ \tau_{2}$) and space ($s_{\mathrm{peak}}$) are provided in Section \ref{sec:basic equations and setting} and \ref{sec:prominence formation}.
We substitute $\bar{Q}_{\mathrm{bg}}(s) = 3.0\times 10^{-4} \ \mathrm{erg \ cm^{-3} \ s^{-1}}$, corresponding to the background heating rate obtained in Section \ref{sec:Coronal heating problem},
and $Q_{\mathrm{L}}$ with the critical value (Case A3, $Q_{\mathrm{local}}=17.0 \ \mathrm{erg \ cm^{-3} \ s^{-1}}$).

Therefore, for a single localized heating event, the required heating rate is approximately $10^{4}$ times higher than the rate derived analytically for the steady heating case.
This value is specific to our numerical setting.
If the value of $s_{\mathrm{peak}}$ increases, condensation will not occur, even with the same amount of heating applied.
Further investigation is needed to understand the general relationship between the required heating amount and the distance between $s_{\mathrm{peak}}$ and the transition region.

Although we discussed heating rates above, the total heating amount is the critical factor for chromospheric evaporation.
Additionally, the numerical setup used in Type A is insufficient for an accurate comparison with the analytical condition \eqref{eq:KL2019}, which assumes steady localized heating applied to both sides of the loop.
To address this discrepancy, we conducted additional simulations with three types of both-sided localized heating: a single heating event similar to Type A in Case B1-4, steady heating in Case C1-5, and steady heating decaying exponentially from the transition regions in Case D1-5 (see Section \ref{Sec:simulation cases}).
The values of $s_{\mathrm{tr}}$ and $\ell_{\mathrm{exp}}$ in Equation \eqref{eq:Localized heating space C} were fixed at 6 Mm and 4 Mm, respectively.
$\tau_{\mathrm{lin}}$ in Equation \eqref{eq:Localized heating time C} was set to 1,500 s.
The results are summarized in Table \ref{table:heating amount}, where the second and third columns present the minimum required heating rate and its integrated value over space and time until condensation occurs.
For Type D, since the spatial distribution contributing to chromospheric evaporation is uncertain, we considered two spatial integration ranges: the region above the photosphere ($0\leq s \leq 10$ Mm) and the transition region ($6\leq s \leq 10$ Mm).
The results show that the total heating amount required for condensation is nearly identical across Type B, C, and D, with at most a variation of a few times, as indicated in the third column of Table \ref{table:heating amount}.
This finding suggests that condensation will occur as long as the total heating flux contributing to chromospheric evaporation remains constant, regardless of the frequency or magnitude of the heating events.

The result of Case D can be directly compared with the analytical condition provided in formula \eqref{eq:KL2019}.
By substituting the value of $Q_{\mathrm{bg}}$ and $Q_{\lambda_{\mathrm{H}}}$, the left-hand side of the formula $\sim 3.6$. On the other hand, the right-hand side lies between $2.0$ and $4.0$, as discussed earlier.
This agreement suggests that our results are consistent with the analytical condition.

Given the above relationship among Type B, C, and D, and the agreement between Type D and the analytical condition, formula \eqref{eq:KL2019} can be extended using the heating amount as follows:
\begin{equation}
    \frac{\int_{s_{\mathrm{tr}}}^{s_{\mathrm{top}}} ds\int_{t_{\mathrm{ini}}}^{t_{\mathrm{cool}}}dt \ Q_{\lambda}}{\int_{s_{\mathrm{tr}}}^{s_{\mathrm{top}}} ds\int_{t_{\mathrm{ini}}}^{t_{\mathrm{cool}}}dt \ Q_{\mathrm{min}}} > 1+\frac{c_{1}}{\Gamma_{\lambda_{\mathrm{H}}}},
\end{equation}
where $t_{\mathrm{cool}}$ is the cooling time.
In Type A, where heating is applied asymmetrically to one side of the loop, the evaporation flow moves toward the opposite side.
This asymmetry increases the required heating amount to more than double compared to Type B, where heating is applied to both sides.
Consequently, this introduces stricter constraints on the conditions necessary for condensation.

\begin{table}[t]
  \caption{The minimum value of the required heating rate and amount}
  \label{table:heating amount}
  \centering
  \begin{tabular}{ccc}
    \hline
    Case & min $Q_{\mathrm{local}}$ [$\mathrm{erg \ cm^{-3}} \ s^{-1}$] & Total [$\mathrm{10^{10} \ erg \ cm^{-2}}$]  \\
    \hline \hline
    B4 & 3.8 & 3.5  \\
    C4 & $5\times10^{-2}$ & 1.8   \\
    D4 & $3\times10^{-3}$ & 1.2-3.1  \\
    \hline
  \end{tabular}
\end{table}

\subsection{Comparison with Previous Studies}\label{Relationship with previous}

Several studies have conducted simulations to investigate condensation triggered by single or transient heating events.
\cite{reep2020electron}, using the HYDRAD code \citep{bradshaw2003self}, explored the potential for coronal rain formation due to a single heating event with electron beams.
They applied short-duration heating (10 and 100 s) in specific parameter spaces but did not observe local condensation.
Instead, runaway cooling seemed to occur throughout the corona.
This runaway cooling likely resulted from the Field number exceeding unity when the coronal density reached $\sim 10^{11} \ \mathrm{cm^{-3}}$ and the temperature fell below $10^{5} \ \mathrm{K}$ (see Figure 4 in their paper).
They suggested that additional mechanisms, which perturb the corona, are necessary to initiate condensation.
In our study, we propose that shock waves serve this role.
Local condensation was observed at multiple locations, even after the loop cooled uniformly to the transition region temperatures.

\cite{kohutova2020self} conducted a self-consistent 3D calculation, finding that condensation was triggered by sudden heating associated with magnetic braiding.
In their study of loop L1, where they observed the complete thermal evolution, they noted that the coronal segment shortened during condensation.
This suggested that the transition region height might increase due to enhanced gas and magnetic pressure gradients caused by condensation, as discussed in Section \ref{sec:overview of prominence formation}.
Since the plasma had already cooled through runaway cooling at this stage, the shortening of the coronal length did not affect condensation, as demonstrated in Section \ref{sec:Field}.
They also investigated the effects of left-right asymmetry in heating, finding that the heating ratio during impulsive heating exceeded 5.
However, this lasted only for about 50 s, and the corona rapidly returned to its original state due to internal flows, thus avoiding incomplete condensation through siphon flow, as seen in cases of steady localized heating \citep[e.g.,][]{mikic2013importance,froment2018occurrence,klimchuk2019role}.
In our case with one-sided heating (Case A1-5), the localized heating lasted less than the cooling time $\tau_{\mathrm{cool}}$.
Condensation occurred after the one-directional flow subsided, similar to \cite{kohutova2020self}'s results.

Many previous studies on condensation phenomena have introduced artificial background heating terms to reproduce the corona \citep[e.g.,][]{antiochos1999dynamic}.
In contrast, \cite{antolin2010coronal} performed calculations that included coronal heating by Alfv$\Acute{\mathrm{e}}$n waves, considering energy dissipation through shock waves.
The shock heating in the corona is episodic in nature, as described by \cite{moriyasu2004nonlinear}.
In their simulation, the maximum vertical velocity amplitude along the loop reached approximately 100 km/s, with an average value around 40 km/s in the corona, which is higher than observed non-thermal velocities.
As presented in Section \ref{sec:Coronal heating problem}, we incorporated additional energy dissipation due to phenomenological Alfv$\Acute{\mathrm{e}}$n wave turbulent cascades, following \cite{shoda2018self}, resulting in vertical velocities that align with observations.
This suggests that our numerical setup provides a more realistic representation of background heating compared to previous studies.

\cite{xia2011formation} reported that once a prominence is formed and left without localized heating, plasma flows into the prominence via a siphon flow due to the lower pressure within the prominence compared to the surrounding corona.
This results in a positive mass change rate for the prominence.
However, our finding reveals a new behavior: the mass change rate is consistently negative, as shown in Figure \ref{fig:prominence formation}(h).
This negative trend is observed across all our studied cases.
We speculate that shock waves colliding with the prominence-corona transition region (PCTR) dissipate energy, heating the prominence and leading to mass drainage.
This result should be further investigated in the future, for example, by using 3D calculations.

In our study, we conducted a parameter survey by varying the magnitude of the localized heating rate.
While this analysis provided valuable insights, other critical factors warrant further investigation, such as the geometry of the loop.
The variation in the cross-sectional area, which was fixed in our study, is intriguing not only from a fluid dynamics perspective \citep[e.g.,][]{mikic2013importance} but also in terms of magnetic wave properties.
Modifying the profile of the expansion rate of the flux tube $f_{\mathrm{ex}}$ in Equations \eqref{eq:expansion factor2} and \eqref{eq:expansion factor} could influence the mode conversion rate.
This behavior should affect the properties of condensation.

While 1D calculations often provide valuable insights into the formation process, extending to multidimensional simulations is essential for accurate comparison with observations.
\cite{zhou2020simulations}, \cite{jervcic2022multi}, and \cite{jervcic2023dynamic} conducted 2D MHD simulations of thread structures in prominences resulting from stochastic heating at the footpoints.
Their studies showed that the magnitude of heating significantly impacts the dynamics of the condensation plasma \citep{jervcic2023dynamic}.
Reproducing inner structures, such as vertical flows and turbulent structures within prominences has mainly been achieved through 3D calculations \citep[e.g.,][]{hillier2012numerical,keppens2015solar,xia2016formation,kaneko2018impact,jenkins2022resolving,donne2024mass}.
Future extensions of our calculations, incorporating heating by MHD waves, may provide further insights into the mass circulation observed by \cite{liu2012first}.

\section{Conclusion}\label{sec:summary}

To investigate the conditions for prominence formation triggered by a single heating event, we performed 1.5D MHD simulations incorporating radiative cooling, thermal conduction, gravity, and a phenomenological heating term.
By applying a boundary driver that mimics the photospheric motion to generate MHD waves, the corona was heated to approximately $T\sim10^{6}$ K.
The loop eventually reached a quasi-steady state.
Localized heating was then applied to the footpoint(s) of the coronal loop, resulting in prominence formation through the ``chromospheric-evaporation condensation" process.

Our findings indicate that Alfv$\Acute{\mathrm{e}}$n waves generated by photospheric motion are converted into shock waves through mode conversion.
As these shock waves pass through the corona, which has become denser due to the localized heating, even denser regions form. In these regions, radiative cooling is further enhanced, leading to condensation.
We conducted a parameter survey varying the magnitudes of the localized heating rate $Q_{\mathrm{local}}$.
It was observed that, for a single heating event, the required heating rate was $\sim 10^{4}$ times greater than in steady heating cases.
However, the total amount of localized heating needed for condensation was similar in both scenarios.
The required heating amount was determined based on the need for temperature decrease due to excess cooling until thermal conduction became inefficient compared to cooling.
Condensation occurred when the Field number $F_{\mathrm{i}}$ approximately exceeded unity.
Furthermore, we extended the condition for thermal non-equilibrium derived by \cite{klimchuk2019role} to a formulation using the heating amount.
This extension makes it applicable even when the localized heating profile does not necessarily decay exponentially and is not steady.

Future studies will involve simulations with different geometries to investigate further the effects of MHD waves and heating mechanisms on condensation.

\vskip\baselineskip

This work was supported by JST SPRING, Grant Number JPMJSP2110.
T. Yokoyama is supported by the JSPS KAKENHI Grant Number JP21H01124, JP20KK0072, and JP21H04492.
T.K. is supported by MEXT/JSPS  KAKENHI Grant No. 20K14519.
Numerical computations were carried out on PC cluster at the Center for Computational Astrophysics, National Astronomical Observatory of Japan.
The authors are grateful to the anonymous referee for improving the manuscript.


\end{document}